   \documentclass[fleqn,usenatbib,usedcolumn]{mnras}
   \usepackage[british]{babel}             % British English hyphenation
   \usepackage{txfonts}                    % Reasonable fonts
   \let\la=\lesssim % for less than similar from txfonts, not \la from mnras.cls
   \usepackage[T1]{fontenc}                % Good font encoding
   \usepackage{graphicx}                   % Including figures
   \hypersetup{pdfauthor={K. Onishi},
               pdftitle={WISDOM Project I: BH Mass Measurement in NGC~3665},
               pdfkeywords={galaxies: individual: NGC~3665 -- galaxies: kinematics and dynamics
  --- galaxies: nuclei --- galaxies: ISM --- galaxies: elliptical and
  lenticular, cD -- galaxies: active},
               bookmarksnumbered=true}
   \volume{{\rm in press}}
\title[WISDOM Project I: BH Mass Measurement in NGC~3665]{WISDOM Project -- I: Black Hole Mass Measurement Using Molecular Gas Kinematics in NGC~3665}

\author[K.\ Onishi et al.]{
Kyoko Onishi,$^{1,2}$\thanks{E-mail: kyoko.onishi@nao.ac.jp}
Satoru Iguchi,$^{1,2}$
Timothy A.\ Davis,$^{3}$
Martin Bureau,$^{4}$
\newauthor
Michele Cappellari,$^{4}$
Marc Sarzi$^{5}$
and Leo Blitz$^{6}$
\\
$^{1}$Department of Astronomical Science, SOKENDAI (The Graduate
University of Advanced Studies), Mitaka, Tokyo 181-8588, Japan\\
$^{2}$National Astronomical Observatory of Japan, Mitaka, Tokyo
181-8588, Japan\\
$^{3}$School of Physics \& Astronomy, Cardiff University, Queens
Buildings, The Parade, Cardiff CF24 3AA, UK\\
$^{4}$Sub-department of Astrophysics, Department of Physics,
University of Oxford, Denys Wilkinson Building, Keble Road, Oxford OX1
3RH, UK\\
$^{5}$Centre for Astrophysics Research, University of Hertfordshire,
Hatfield, Herts AL1 9AB, UK\\
$^{6}$Department of Astronomy, University of California, Berkeley,
California 94720, USA
}

\date{Accepted for publication in MNRAS}

\pubyear{2017}

\begin{document}
\label{firstpage}
\pagerange{\pageref{firstpage}--\pageref{lastpage}}
\maketitle

%It should be a single paragraph not more than 250 words (200 words for Letters).
%No references should appear in the abstract.
\begin{abstract}
  As a part of the mm-Wave Interferometric Survey of Dark Object Masses (WISDOM) project, we present an estimate of the mass of the supermassive black hole (SMBH) in the nearby fast-rotator early-type galaxy NGC~3665. We obtained Combined Array for Research in Millimeter Astronomy (CARMA) B and C array observations of the $^{12}$CO$(J=2-1)$ emission line with a combined angular resolution of $0\farcs59$. We analysed and modelled the three-dimensional molecular gas kinematics, obtaining a best-fit SMBH mass $M_{\rm BH}=5.75^{+1.49}_{-1.18} \times 10^{8}$~$M_{\odot}$, a mass-to-light ratio at $H$-band $(M/L)_{H}=1.45\pm0.04$~$(M/L)_{\odot, H}$, and other parameters describing the geometry of the molecular gas disc (statistical errors, all at $3\sigma$ confidence). 
We estimate the systematic uncertainties on the stellar $M/L$ to be $\approx0.2$~$(M/L)_{\odot, H}$, and on the SMBH mass to be $\approx0.4\times10^{8}$~$M_{\odot}$.
The measured SMBH mass is consistent with that estimated from the latest correlations with galaxy properties. Following our older works, we also analysed and modelled the kinematics using only the major-axis position-velocity diagram, and conclude that the two methods are consistent.
\end{abstract}


%%%%%%%%%%%%%%%%%%%%%%%%%%%%%%%%%%%%%%%%%%%%%%%%%%

%%%%%%%%%%%%%%%%% BODY OF PAPER %%%%%%%%%%%%%%%%%%

%
% Section --- Introduction
%
\section{Introduction}
\label{sec:intro}

Substantial improvements in observational capabilities over the past two decades have allowed to measure the mass of supermassive black holes (SMBHs) in nearby galaxies using a few different methods across a range of wavelengths, revealing that SMBHs are ubiquitous and most likely present in every stellar spheroid (see, e.g.,
\citealt{2013ARA&A..51..511K} for a review). Black holes are now recognized to play a fundamental but as yet not fully understood role in the growth and evolution of galaxies.

As the angular resolution of observations increases, so does the number of reliable SMBH mass measurements, and several correlations have now emerged between SMBH mass and host galaxy properties such as bulge mass \citep[e.g.][]{1998AJ....115.2285M, 2003ApJ...589L..21M, 2012MNRAS.419.2497B, 2013ARA&A..51..511K}, total luminosity \citep[e.g.][]{2014ApJ...780...69L} and bulge velocity dispersion \citep[e.g.][]{2000ApJ...539L...9F, 2009ApJ...698..198G, 2013ApJ...764..184M, 2013ARA&A..51..511K}. This last so-called $M_{\rm BH}-\sigma$ relation is the tightest, but all empirical correlations suggest that galaxy evolution and SMBH growth are closely connected. Numerical simulations reproduce these correlations and provide suggestive evidence that this co-evolution likely involves self-regulation mechanisms (e.g.\ AGN feedback; \citealt{1998A&A...331L...1S, 2008ApJ...676...33D, 2014MNRAS.437.1456B}).

Another fact supporting the co-evolution paradigm is that as the number of dynamically-measured SMBH masses increases, different relationships for different types of galaxies are starting to emerge. For example, \citet{2013ApJ...764..184M} found that different relations best fit samples of early- and late-type galaxies, the SMBH
mass being typically two times larger at a given velocity dispersion for early-type galaxies. The current empirical correlations are however based on only $\approx80$ objects, not enough to reliably divide objects and contrast the relations of different galaxy samples. This shortcoming is compounded by the fact that almost $70\%$ of these objects are early-type galaxies, a problem arising from the very limited number of methods available to measure SMBH masses, and the fact that each method is biased toward objects with certain characteristics. We discuss this problem below, and the need for more measuring techniques applicable across a broader range of galaxy types.

Dynamical SMBH mass measurements have so far relied on only a handful of methods. Stellar dynamical measurements have the obvious advantage that every galaxy has a substantial population of stars moving exclusively under the influence of gravity, and stars are easily probed at optical and near-infrared (NIR) wavelengths. The modelling methods are well established and applicable to many galaxies; $\gtrsim60\%$ of current SMBH mass measurements rely on stellar dynamics \citep[see, e.g.,][]{1988ApJ...324..701D, 1998AJ....116.2220V, 2002ApJ...578..787C, 2003ApJ...583...92G, 2005ApJ...628..137V, 2011Natur.480..215M, 2012Natur.491..729V, 2013AJ....146...45R}. Nevertheless, the dynamical modelling tools are relatively complex and often restricted to axisymmetric objects, dust can easily perturb measurements, and relatively high spectral resolution is required to probe the higher order moments of the line-of-sight velocity distributions. These constraints result in
samples heavily biased toward early-type galaxies.

Ionised gas dynamics is also used to measure SMBH masses, and is typically probed through nebular emission lines at optical/NIR wavelengths, that are present in most galaxies \citep[see, e.g.,][]{1996ApJ...470..444F, 1997ApJ...489..579M, 1998AJ....116.2220V, 2002ApJ...578..787C, 2007ApJ...671.1329N, 2008A&A...479..355D, 2013ApJ...770...86W}. Challenges for this method however include non-gravitational forces (e.g.\ shocks) and significant turbulent motions superimposed on (quasi-)circular motion. Observations with multiple slits are often unable to fully characterise the kinematics and overcome these challenges, but significant improvements have been made with integral-field spectroscopic (IFS) observations \citep[see, e.g.,][]{2007ApJ...671.1329N, 2016ApJ...817....2W}.

The dynamics of accretion discs containing megamasers offers another way forward to measure SMBH masses \citep[see, e.g.,][]{1995Natur.373..127M, 2011ApJ...727...20K}. The precision is nearly on par with that in our own Milky Way \citep[e.g.][]{2008ApJ...689.1044G}, occasionally allowing to positively rule in SMBHs (as opposed to other potential compact dark objects), but the rarity of appropriate megamaser systems ($\sim1\%$ of objects searched; see \citealt{2005ARA&A..43..625L, 2016ApJ...819...11V}) makes this at best a niche method.  Another limitation of this method is that megamasers have only been detected in Seyfert 2 and low-ionisation nuclear emission-line region (LINER) galaxies so far, containing SMBHs of relatively low masses. The reason for this bias is unknown, but it again suggests a difficulty to build up large samples.

More recently, the dynamics of molecular gas probed at mm/sub-mm wavelengths has emerged as a very promising method \citep[see, e.g.,][]{2013Natur.494..328D, 2015ApJ...806...39O, 2016ApJ...823...51B, 2016ApJ...822L..28B}, in particular because of the exquisite angular resolution and sensitivity afforded by the Atamaca Large Millimeter/sub-millimeter Array (ALMA). 
As for ionised gas, molecular gas can in principle be affected by non-gravitational forces and turbulence \citep[e.g.][]{2003A&A...407..485G, 2015A&A...583A.104S}, although in practice (giant) molecular clouds tend to move ballistically over much of their orbits, and molecular gas observations are unaffected by dust. 
Of course, some galaxies have no molecular gas detected (as is the case for many early-type galaxies; \citealt{2011MNRAS.414..940Y}), and others none in their very centre \citep[][]{2003ApJS..145..259H, 2013MNRAS.432.1796A}, but this method has enormous potential. 
For instance, \citet{2014MNRAS.443..911D} showed that the SMBH mass of $\approx35,000$ local galaxies could be measured with ALMA when it reaches full capability, and the sphere of influence (SOI; $R_{\rm SOI}\equiv GM_{\rm BH}/\sigma^{2}$, where $G$ is the gravitational constant) of the largest SMBHs ($M_{\rm BH}\geq 10^{8.5}$~$M_{\odot}$) is spatially resolvable across the whole of cosmic time. Needless to say, the molecular gas method also has the potential to redress the current bias against late-type galaxies in $M_{\rm BH}-\sigma$ studies (now only $\approx30\%$ of the sample).

Both \citet{2013Natur.494..328D} and \citet{2015ApJ...806...39O} derived the SMBH mass without directly detecting the Keplerian motion at the galaxy centre. This was due to the particular molecular gas disc morphology (in the case of \citealt{2013Natur.494..328D}) and severe beam smearing (in the case of \citealt{2015ApJ...806...39O}). 
The SMBH masses were constrained to good accuracy (e.g. 20\% error for \citealt{2015ApJ...806...39O}), although higher-angular resolution would have allowed to further constrain the measured values.
\citet{2016ApJ...823...51B} detected the Keplerian upturn at the centre of the nearby galaxy NGC~1332, but the SMBH mass was not very well constrained due to the SOI resolved along the disc's major axis but unresolved along its minor axis. Higher-angular resolution observation in \citet{2016ApJ...822L..28B} constrained the SMBH mass to $10\%$ accuracy.

Building on small pilot projects \citep{2013Natur.494..328D, 2015ApJ...806...39O}, we have recently started the mm-Wave Interferometric Survey of Dark Object Masses (WISDOM) project. This project aims to benchmark and test the molecular gas dynamics method, develop tools and best practice, and exploit the growing power of ALMA to better populate and thus constrain SMBH -- galaxy scaling relations. This paper is the first of a series, and introduces the tools and fitting procedures developed so far, in the context of estimating the SMBH mass in the nearby fast-rotator early-type galaxy NGC~3665. We use the method initially employed by \citet{2013Natur.494..328D}, but further extend it to exploit the full three-dimensional data cubes.

This paper is stuctured as follows. The galaxy selection, observations, and data reduction and analysis are described in Section~\ref{sec:obs}. The SMBH mass measurement method is explained in Section~\ref{sec:model}. Section~\ref{sec:dis} contains a discussion of galaxy morphology, AGN activity, possible sources of error on the SMBH mass, and a comparison of the SMBH masses derived using different fitting methods. Our main results are summarized in Section~\ref{sec:conc}.

%
% Section --- NGC3665 properties and observations
%
\section{Data}
\label{sec:obs}

% Subsection --- NGC3665
\subsection{NGC~3665}
\label{subsec:obs_n3665}

NGC~3665 is a nearby fast-rotator early-type galaxy \citep{2011MNRAS.414..888E}. We adopt a distance of $34.7$~Mpc, estimated from the Tully-Fisher relation by \citet{2007A&A...465...71T}, yielding a scale of $\approx167$~pc~arcsec$^{-1}$. This distance is in good agreement with the estimate of $33.1$~Mpc from its recession velocity\footnote{The choice of the distance does not influence our conclusions but sets the scale of our models in physical units. Specifically, lengths and masses scale as $D$, while $M/L$s scale as $D^{-1}$.} by \citet{2011MNRAS.413..813C}. Basic properties of the galaxy are summarised in Table~\ref{table:n3665properties}. Twin jets emanating from the galaxy nucleus were observed with arcsecond resolution at $1.5$~GHz using the Very Large Array (VLA; \citealt{1986A&AS...64..135P}), while a point-like structure was observed with $2$ milli-arcsecond resolution at $5$~GHz using the Very Long Baseline Array (VLBA; \citealt{2009A&A...505..509L}), both proving the existence of a central SMBH. We adopt the position of this point-like structure as the galaxy centre: RA=11:24:43.624, Dec=38:45:46.278 \citep{2009A&A...505..509L}.

% Table: NGC3665 properties
\begin{table}
  \begin{center}
    \caption{NGC~3665 properties.}
    \label{table:n3665properties}
    \begin{tabular}{lrr} \hline
      Parameter& Value & Reference \\ 
      \hline
      Morphology & fast rotator early type & 1 \\
      Position &  & 2 \\
      \ \ \ RA (J2000.0) & $11^{\rm h}24^{\rm m}43^{\rm s}\!\!.624$ & \\
      \ \ \ DEC (J2000.0) & $38\degr45\arcmin46\farcs278$ & \\
      Systemic velocity (km~s$^{-1}$) & 2069 & 3 \\
      Position angle ($\degr$) & 26 & 4 \\
      Inclination angle ($\degr$) & $69.9^{+0.51}_{-0.39}$ & 4 \\
      Distance (Mpc) & $34.7 \pm 6.8$ & 5,6 \\
      Linear scale (pc~arcsec$^{-1}$) & $167 \pm 33$ & 5 \\
      \hline
    \end{tabular}
  \end{center}
  References: (1) \citet{2011MNRAS.414..888E}; (2)
  \citet{2009A&A...505..509L}; (3) \citet{2011MNRAS.413..813C}; (4)
  this work; (5) \citet{2007A&A...465...71T}; (6) NASA/IPAC
  Extragalactic Database (\small http://ned.ipac.caltech.edu/).
\end{table}

\begin{figure*}
 \includegraphics[width=17.0cm]{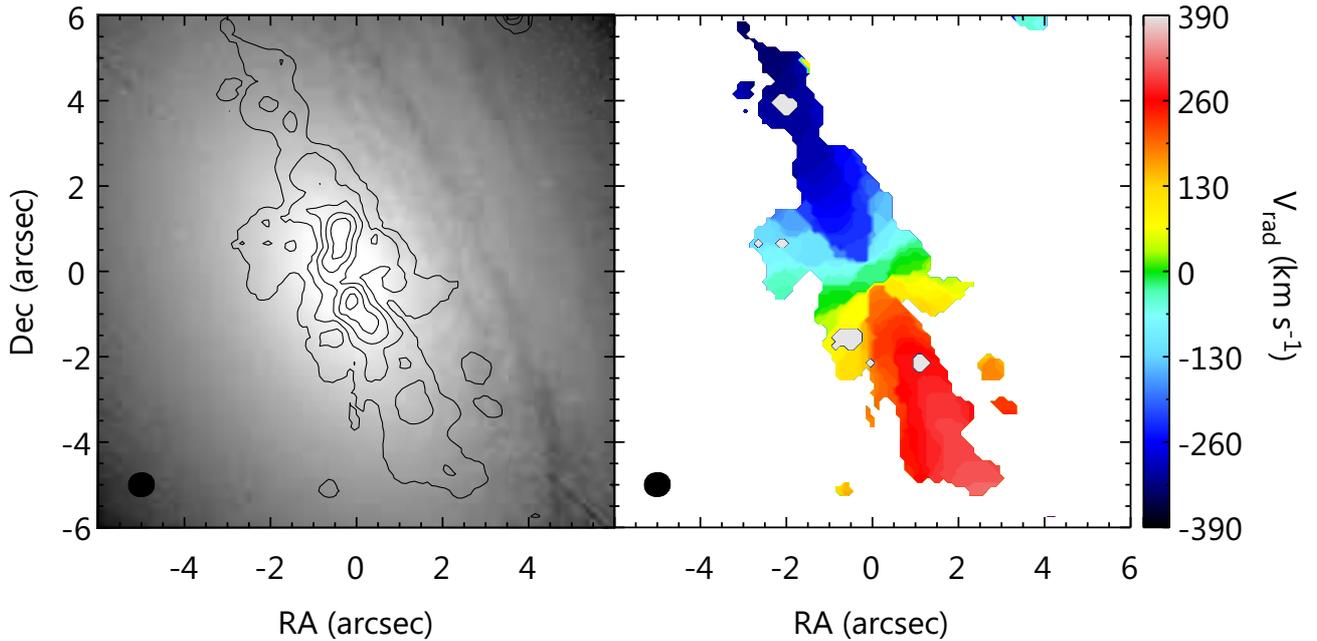}
  \caption[Integrated intensity and mean velocity maps.]{Left panel:
    Integrated intensity map of CO$(J=2-1)$ in NGC~3665 (contours),
    overlaid on the \textit{HST} $H$-band (NICMOS F160W) unsharp-masked image.
    Contour levels are $0.3$, $0.9$, $1.5$, $2.1$ and
    $2.7$~Jy~beam$^{-1}$~km~s$^{-1}$, while the peak flux is
    $3.3$~Jy~beam$^{-1}$~km~s$^{-1}$. Right panel:
    Intensity weighted (mean) velocity map of CO$(J=2-1)$. The map
    extent is set to the lowest contour level of the integrated
    intensity map. The synthesised beam ($0\farcs60\times0\farcs56$ at
    a position angle of $-81\degr$) is shown in the bottom-left
    corner of each panel. }
  \label{fig:mom01}
\end{figure*}

\textit{Hubble Space Telescope} (\textit{HST}) observations in $H$-band (NICMOS F160W; see the left panel of Figure~\ref{fig:mom01}) show prominent and regular dust lanes circling the galaxy centre, suggesting a gaseous disc-like structure extending to a radius of at least $8\arcsec$ from the nucleus. Previous CARMA observations revealed a regularly rotating molecular gas disc of the same radius \citep{2013MNRAS.432.1796A}. We therefore expect a central SMBH surrounded by a relaxed molecular gas disc, ideal to measure the SMBH mass using molecular gas kinematics.

% Figure: NGC3665 MOM0 and MOM1 maps

NGC~3665 does not have a SMBH mass measurement yet, but has an effective stellar velocity dispersion of $\sigma_{\rm e}=216 \pm 10$~km~s$^{-1}$, reliably measured through integral-field spectroscopy \citep{2013MNRAS.432.1709C}. Adopting the $M_{\rm BH}-\sigma$ relation of \citet{2013ApJ...764..184M}, this suggest a SMBH mass of $\approx3\times10^{8}$~$M_{\odot}$ and thus a SMBH SOI of $\approx30$~pc or $\approx0\farcs2$, that is 3 times smaller than the synthesised beam size of the current observations (see Section~\ref{subsec:obsdr}).

% Subsection --- Observations
\subsection{Observations and Data Reduction}
\label{subsec:obsdr}

NGC~3665 was observed at the wavelength of the $^{12}$CO$(J=2-1)$ line using CARMA in the B (baselines $63$--$947$~m) and C (baselines $26$--$370$~m) arrays. Observations were carried out from April 11th to 21st 2013 (B array) and from November 26th to December 14th 2013 (C array). Total on-source time was 1610.6~min in the B array and 290~min in the C array. The receivers were tuned to cover the frequency range $214.404$--$215.373$~GHz and $228.576$--$229.545$~GHz in the lower (LSB) and upper (USB) sidebands, respectively, with $4$ spectral windows per range. With $31$ channel per spectral window, the frequency resolution was thus $7.812$~MHz per channel. The field of view (full width at half maximum, FWHM, of the primary beam) at these frequencies was $31\arcsec$ for the $10$-m antennae. The observational parameters are summarized in Table~\ref{table:obsproperties}.

% Table: CARMA observing parameters
\begin{table}
  \begin{center}
    \caption{CARMA observation parameters.}
    \label{table:obsproperties}
    \begin{tabular}{lcc} \hline
      Parameter & B array & C array \\ 
      \hline
      Date & Nov-Dec 2013 & Apr 2013 \\
      On-source time (min) & $1610.6$ & $290$ \\
      Phase center: &  &  \\
      \ \ \ RA(J2000.0) & \multicolumn{2}{c}{$11^{\rm h}24^{\rm m}43^{\rm s}\!\!.6$}\\
      \ \ \ DEC(J2000.0) & \multicolumn{2}{c}{$38\degr45\arcmin46\farcs278$}\\
      Primary beam & \multicolumn{2}{c}{$31\arcsec$}\\
      & & \\
      \hline
      & LSB & USB \\
      \hline
      Frequency coverage (GHz) & $214.404$--$215.373$ & $228.576$--$229.545$ \\
      Velocity resolution (km~s$^{-1}$) & $10.9$ & $10.2$ \\
      \hline
    \end{tabular}
  \end{center}
\end{table}

We followed the data reduction and analysis method described in \citet{2013MNRAS.432.1796A} using the Multichannel Image Reconstruction Image Analysis and Display (MIRIAD) package \citep{1995ASPC...77..433S}. The visibility data were first edited and calibrated using Mars, MWC349 and 3C273 as flux calibrators and 1153+495 as phase calibrator. The bandpass calibrator was 3C279 or 3C273.
We then subtracted the continuum by assuming the CO emission to be present from $-450$ to $450$~km~s$^{-1}$ with respect to the galaxy systemic velocity of $2069$~km~s$^{-1}$ \citep[][]{2011MNRAS.413..813C}, and subtracting a linear fit to the line-free channels. The resulting cube was cleaned using the MIRIAD task \textsc{MOSSDI2} with a threshold of 1.5 times the rms noise, measured in line-free regions of the cube. For imaging, we set the robustness parameter to $0.5$, yielding a synthesised beam FWHM of $0\farcs60\times0\farcs56$ ($\approx100$~pc $\times$ $93$~pc) at a position angle of $-81\degr$, that was properly sampled with $0\farcs2\times0\farcs2$ pixels. This provides a compromise between angular resolution and sensitivity, and the rotational motion of the molecular gas in the galaxy nucleus is then clearly detected. We did not bin the velocity channels, yielding a velocity resolution of $7.8$~MHz or $\approx10$~km~s$^{-1}$ per channel, and the average rms noise per channel was $4.1$~mJy~beam$^{-1}$. Molecular gas emission was finally detected from $-370$ to $380$~km~s$^{-1}$. 

Integrated intensity (moment~0) and intensity-weighted (mean) velocity (moment~1) maps can be created directly from this cube, but as most of the cube is devoid of emission, the resulting maps are of poor quality. Instead, we optimised the moments by first Hanning-smoothing the data cube in velocity and then smoothing it spatially with a Gaussian of FWHM equal to that of the synthesised beam. A mask was then created by selecting all pixels in the smoothed data cube above a threshold of $0.75$ times the rms noise in each channel. The adopted integrated intensity and mean velocity maps, shown in Figure~\ref{fig:mom01}, were then created by calculating the moments of the original unsmoothed data cube within the mask region only.

The integrated intensity map reveals a centrally-concentrated molecular gas distribution, rapidly decreasing with radius. However, the inner $\approx2\arcsec$ clearly show two separate concentrations on either side of the nucleus (see Figure~\ref{fig:mom01}, left panel). This suggests a void in the very centre of the galaxy, where the intensity may remain above zero simply due to the angular extent of the synthesised beam. This central hole is confirmed by our modelling and further discussed in Section~\ref{subsec:morph}. The mean velocity map (Figure~\ref{fig:mom01}, right panel) reveals very regular disc-like rotation with a total velocity width of $\approx750$~km~s$^{-1}$ and no evidence of any significant non-circular motion, warp or kinematic twist.

Interestingly, the molecular gas distribution and dust lanes revealed by \textit{HST} are not very well associated with each other (see Figure~\ref{fig:mom01}, left panel). The kinematic major axis of the molecular gas and the major axis of the dust lanes are however aligned, suggesting that the molecular gas disc and dust are nevertheless in the same plane. Both also align well with the large-scale photometric major axis, suggesting that this plane is also the large-scale equatorial plane of the galaxy. The differing distributions of the molecular gas and dust may thus be due to missing flux in the interferometric data, due to the lack of truly short baselines. Indeed, comparing the flux from our observations integrated over the CO$(2-1)$ Institut de Radioastronomie Millimetrique (IRAM) 30-m telescope beam ($42.3$~Jy~km~s$^{-1}$) to an actual IRAM 30-m integrated flux measurement ($67.1$~Jy~km~s$^{-1}$; \citealt{2011MNRAS.414..940Y}), our CARMA high-resolution observations may be resolving out $\approx40\%$ of the flux in extended structures.

%
% Section --- SMBH mass estimate
%

\section{Supermassive Black Hole Mass Estimate}
\label{sec:model}

In this section, we describe the procedures employed to measure the SMBH mass in NGC~3665 and state our results. To summarise, we modelled the three-dimensional stellar mass distribution of the galaxy by deprojecting a two-dimensional model of the observed surface brightness and assuming a constant mass-to-light ratio $M/L$. The circular velocity curve arising from this mass model and a putative SMBH was then fed into a code simulating the resulting data cube, taking into account the molecular gas distribution and instrumental effects. The SMBH mass was then determined by simply comparing a range of models to the observations.

% Subsection --- Velocity Model
\subsection{Velocity Model}
\label{subsec:model_lumdis}

The galaxy mass distribution is assumed to be the sum of a central SMBH and the large-scale stellar body of the galaxy. The SMBH is treated as a point mass whose mass is free. For the stars, we combine \textit{HST} (NICMOS F160W) and Two Micron All-Sky Survey (2MASS) $H$-band images, allowing to accurately trace the stellar surface brightness to a radius of $\approx40\arcsec$. We adopt the Multi Gaussian Expansion (MGE) method \citep{1994A&A...285..723E,   2002ApJ...578..787C} and fit this two-dimensional image with a sum of Gaussians (\textsc{MGE\_FIT\_SECTORS} procedure\footnote{Available from http://purl.org/cappellari/software} of \citet{2002MNRAS.333..400C}). Given an inclination each Gaussian can be deprojected analytically, and the three-dimensional light distribution of the model can thus be trivially reconstructed.  Here all the Gaussians are constrained to have the same position angle and inclination, resulting in an axisymmetric light model. 

\begin{figure}
 \includegraphics[width=\columnwidth]{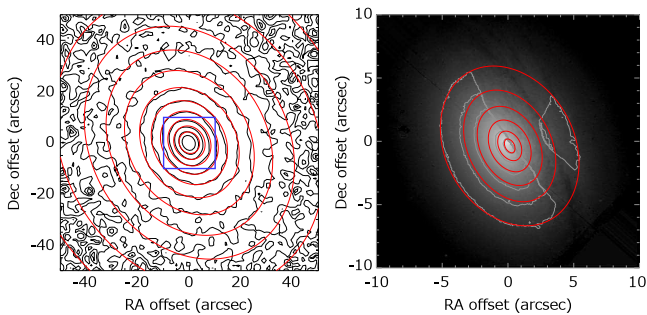}
  \caption[\textit{HST} $H$-band image and MGE model.]{Left panel: 2MASS $H$-band image (black contours)
    with our MGE model overlaid (red contours; Table~\ref{table:MGE}). Right panel: Our MGE model (red contours) of the central $10\arcsec$ (blue box in the left panel), overlaid on the \textit{HST} $H$-band (NICMOS F160W) image (grey contours and grey scale image).
    The masked region (due to dust) is identified without the grey contours. }
  \label{fig:MGE}
\end{figure}

The point spread function (PSF) of each image is also fit with a sum of (circular) Gaussians and used as input during the MGE fit to obtain a deconvolved light model of the galaxy. We use the Tiny~Tim package (version~6.3) developed by \citet{2011SPIE.8127E..0JK} to measure the \textit{HST} NICMOS PSF of the F160W filter. The FWHM of the 2MASS $H$-band PSF was assumed to be $2\farcs8$\footnote{http://spider.ipac.caltech.edu/staff/roc/2mass/seeing/seesum.html}. The prominent dust lane seen in the \textit{HST} image is masked to mitigate the effects of dust obscuration, and the region outside a radius of $6''$ in the \textit{HST} image is ignored and overwritten with the 2MASS image. Figure~\ref{fig:MGE} shows a comparison of the best-fitting MGE model and the observed surface brightness distribution of the galaxy, from both 2MASS and \textit{HST}. We follow the photometric calculation described in Section~5.2 of \citet{2009nicm.book.....T} to convert the flux units from counts~pixel$^{-1}$~second$^{-1}$ to $L_{\odot}$~pc$^{-2}$, and the resulting MGE parameters are listed in Table~\ref{table:MGE}. We adopt an $H$-band solar Vega magnitude $M_{\odot, H}=3.32$ from Table~1 of \citet{2007AJ....133..734B}.　

% Figure: MGE model

% Table: MGE components
\begin{table}
  \begin{center}
    \caption{MGE components of the \textit{HST} NICMOS F160W and 2MASS $H$-band images.}
    \label{table:MGE}
    \begin{tabular}{lrrr}
      \hline
      $j$ & $I_{j}$\phantom{2222} & $\sigma_{j}$\phantom{2.} & $q_{j}\phantom{2}$ \\
      & ($L_{\odot, H}$~pc$^{-2}$) & (arcsec) & \\
      \hline
      1.......... & $25551.5$\phantom{22} & $0.227$ & $0.515$ \\
      2.......... & $21118.8$\phantom{22} & $0.661$ & $0.608$ \\
      3.......... & $7436.97$\phantom{2} & $1.31$\phantom{2} & $0.887$ \\
      4.......... & $12016.7$\phantom{22} & $2.17$\phantom{2} & $0.576$ \\
      5.......... & $5862.67$\phantom{2} & $4.76$\phantom{2} & $0.837$ \\
      6.......... & $741.344$ & $11.3$\phantom{22} & $0.441$ \\
      7.......... & $807.669$ & $19.2$\phantom{22} & $0.780$ \\
      8.......... & $212.118$ & $48.6$\phantom{22} & $0.821$ \\
      \hline
    \end{tabular}
  \end{center}
\end{table}

The gravitational potential of the galaxy is calculated from the summation of the MGE model components multiplied by a constant $M/L$ and the SMBH modelled as a point like mass, by following the equation in \citet{2002ApJ...578..787C}. The circular velocity curve in the equatorial plane is then calculated from this by using the ; \textsc{mge\_circular\_velocity} procedure within the Jeans Axisymmetric Modelling (JAM) package\footnote{http://purl.org/cappellari/software} (\citealt{2008MNRAS.390...71C}). 

Our assumption of a spatially constant $M/L$ should be treated with caution given that the molecular gas disc (and potentially associated star formation) extends to the nucleus of NGC~3665 \citep{2014MNRAS.444.3427D}. The effect of any uniform star formation or young stellar population in the region of interest is implicitly subsumed into our adopted $M/L$, but a steep gradient could be problematic. 
However, neglecting centrally-concentrated star formation, and thus a decreasing $M/L$ with decreasing radius, effectively overestimates the stellar contribution to the total mass in the very centre. Our SMBH mass estimates are thus conservative.
The effects of a potentially varying $M/L$ will be further explored in \citet{2017MNRAS.464..453D}.

% Subsection --- Kinematic modelling
\subsection{Data Cube Model}
\label{subsec:model_velcalc}

Given the circular velocity curve obtained from the MGE formalism described above (Section~\ref{subsec:model_lumdis}), and the adopted molecular gas disc inclination (which is the same as that used to de-project our MGE models), we generate a model data cube using the Kinematic Molecular Simulation (KinMS) code of \citet{2013Natur.494..328D}\footnote{https://github.com/TimothyADavis/KinMS}. 
This assumes circular motions and a spatially uniform (but free) gas velocity dispersion. Instrumental effects such as beam-smearing and spatial and velocity binning are all taken into account by KinMS. For the properties of the molecular gas disc, we further assume that it has an exponential surface brightness profile with a void in the centre. The three free parameters describing the molecular gas distribution are thus its surface brightness scaling factor, radial scale radius and the radius of the central hole. The other free parameters required to fully describe a model (or, rather, to allow its comparison to real data) are the kinematic centre and position angle of the molecular gas disc in the plane of the sky and the galaxy systemic velocity. The total list of $11$ free parameters is given in Table~\ref{table:params}.

% Subsection --- Model fit
\subsection{Model Fitting}
\label{subsec:fitting}

We use Bayesian analysis techniques to estimate the best-fit set of model parameters from our data cube, including the SMBH mass. Specifically, we utilize a Markov chain Monte Carlo (MCMC) method with Gibbs sampling to explore the parameter space. The number of iterations is set to $10^{6}$, and the first $5\times10^{4}$ iterations are ignored as a burn-in phase. The method will be fully described in \citet{2016Davisetal.MNRAS.submitted}, but we provide a short summary here. The aim is to obtain the posterior distribution of the $11$ model parameters: SMBH mass, stellar $M/L$, and the molecular gas disc kinematic centre, inclination, position angle, systemic velocity, velocity dispersion, integrated flux (CO surface brightness scaling factor), radial scalelength and void radius. 

The region of the cube used for fitting covers the entire CO emitting region, and is defined to be $13\farcs0 \times 13\farcs0$, the centre coinciding with the core position observed by \citet{2009A&A...505..509L}. Velocity channels are from $-380$ to $380$~km~s$^{-1}$ with respect to the systemic velocity of $2069$~km~s$^{-1}$.

We use a logarithmic likelihood function based on the $\chi^{2}$ distribution, calculated by comparing the CO distribution in each channel of the data cube with that in the model. As our data are approximately Nyquist sampled spatially, the synthesised beam induces strong correlations between neighbouring pixels in the data cube. The likelihood function we use, $\exp(-\chi^{2}/2)$, takes this into account by including the full covariance matrix when calculating the $\chi^{2}$. As the condition number of the covariance matrix itself is large, we do not invert it directly to calculate the likelihood, but instead introduce a modified Cholesky factorization step to avoid loss of numerical precision when calculating the inverse. The observational error on the flux in each pixel is set to the rms noise of the data cube evaluated in the central regions, in channels where no emission is detected.

\begin{figure*}
    \includegraphics[width=17.0cm]{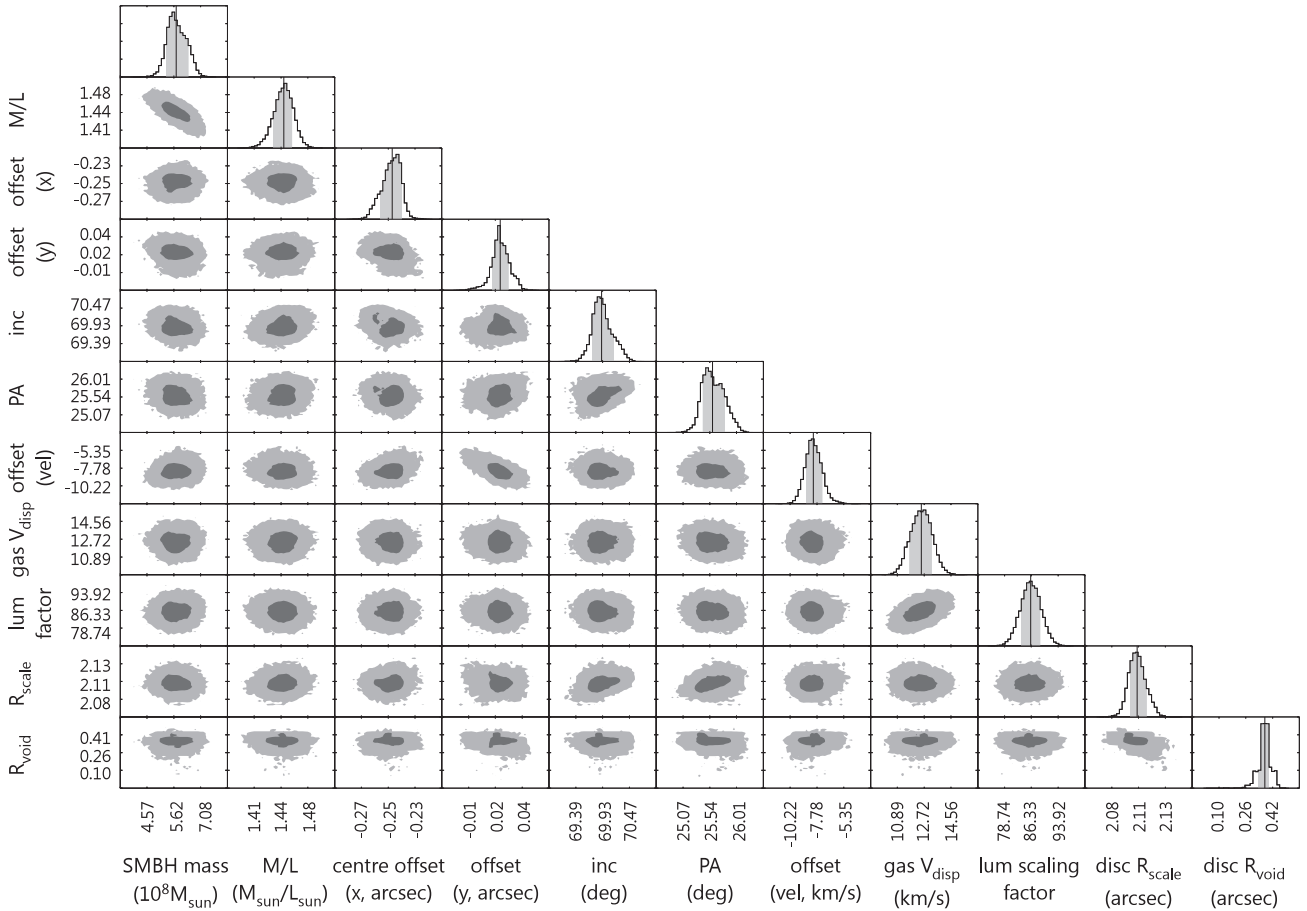}
    \caption[posterior histograms and likelihood contours.]{Histograms showing the
      posterior distribution of each model parameter, with the
      $68.3\%$ ($1\sigma$) confidence interval shaded in
      grey. Greyscales show the likelihood distribution of every pair
      of parameters. Regions of parameter space within the $3\sigma$
      confidence level are coloured in pale grey while regions within
      $1\sigma$ are coloured in dark grey. Some pairs of parameters show a
      correlation (e.g.\ SMBH mass and $M/L_{H}$), but they are still tightly constrained.
      The vertical lines in the histograms show the best-fit value of each parameter.
      See Table~\ref{table:params} for a quantitative
      listing of the uncertainties.}
    \label{fig:histocontours}
  \end{figure*}

\begin{figure*}
   \includegraphics[width=17.0cm]{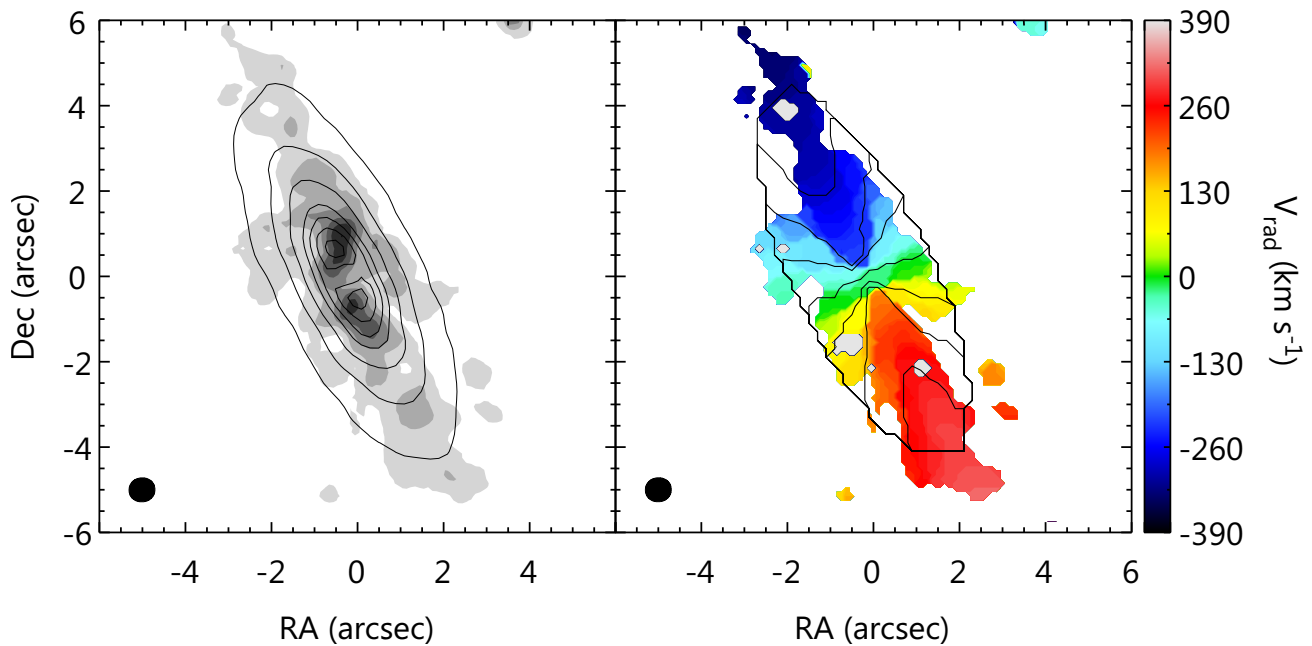}
  \caption[Observed and best-fit model moment~0 and moment~1
  maps.]{Left panel: Integrated intensity (moment~0) map of the CARMA
    observations (greyscale), overlaid with that of the best-fit model
    (contours). Contours are set to be $1/12$, $2/12$, $3/12$, $5/12$, $7/12$, $9/12$ and $11/12$ of the peak. 
    Right panel:
    Intensity-weighted (mean) velocity (moment~1) map of the the CARMA
    observations (colourscale), overlaid with that of the best-fit
    model (contours). Contours are spaced by $130$~km~s$^{-1}$ from $-390$ to $390$~km~s$^{-1}$. The synthesised beam ($0\farcs60\times0\farcs56$ at a position angle
    of $-81\degr$) is shown in the bottom-left corner of each panel.}
  \label{fig:mom01compare}
\end{figure*}

\begin{figure*}
   \includegraphics[width=17.0cm]{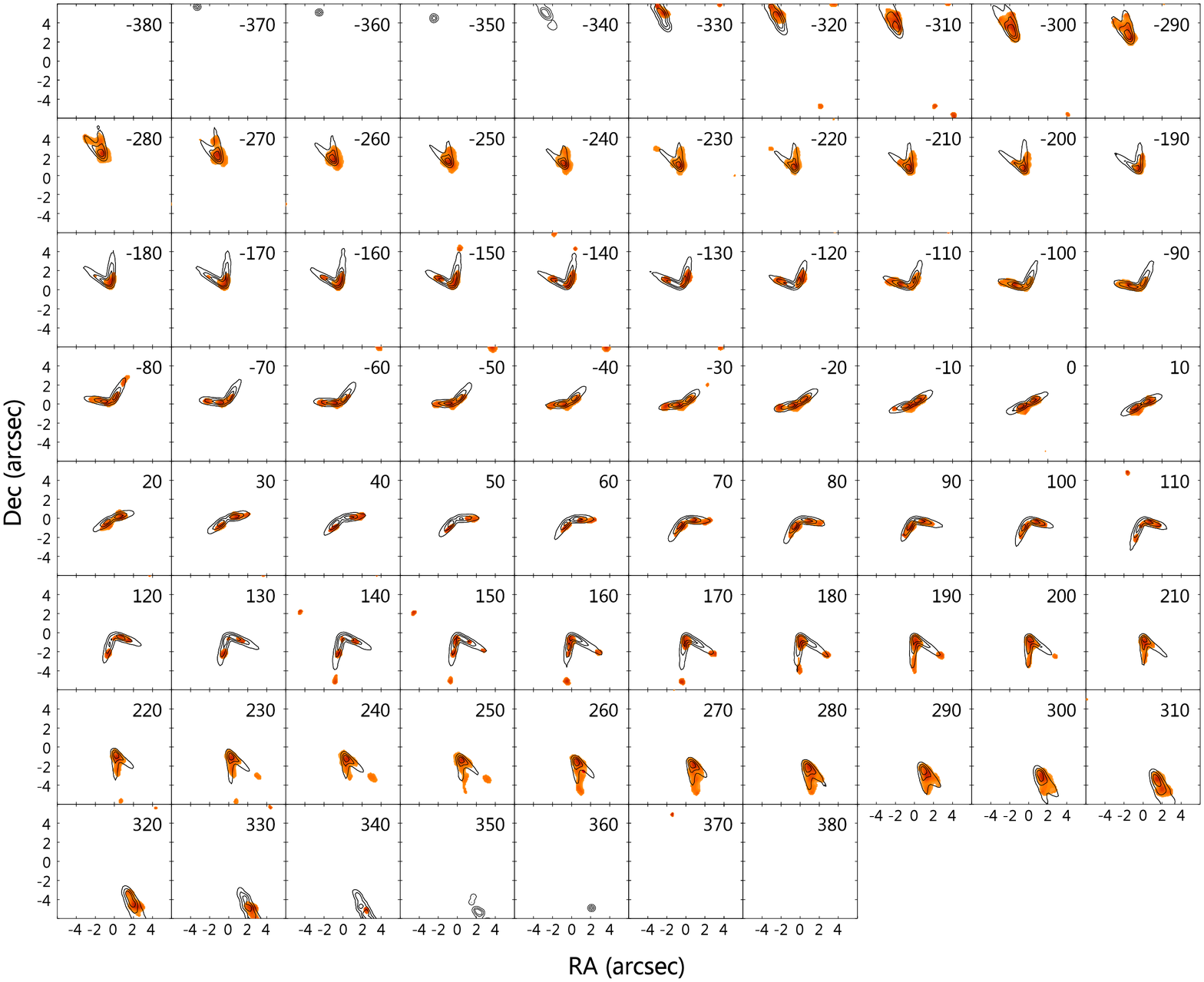}
  \caption[Observed and best-fit model channel maps.]{Channel maps of
    the CARMA observations (colour scale), overlaid with those of the
    best-fit model (black contours). The velocity of each channel in
    km~s$^{-1}$ is indicated in the top-right corner of each panel. Contours of the observations are set to be $3/15$, $5/15$, $7/15$, $9/15$, $11/15$ and $13/15$ of the peak intensity observed in each channel. Models are plotted with contours set to be $3/15$, $7/15$ and $11/15$ of the peak intensity of the model in each channel. }
  \label{fig:channelmaps}
\end{figure*}

We use flat priors for all the fitted parameters, within certain ranges. The prior distributions used and the posterior distributions returned are summarised in Table~\ref{table:params}. The posterior distributions are also shown with greyscales in Figure~\ref{fig:histocontours}. A comparison of the data and best-fit model moment~0 and moment~1 maps is shown in Figure~\ref{fig:mom01compare}. An analogous comparison of the channel maps is shown in Figure~\ref{fig:channelmaps}.

% Figure: Chi-square plots
%\begin{landscape}
%\end{landscape}

% Figure: Model moment 0 and 1 comparison

% Figure: Channel map comparison

% Sub-section --- Model results
\subsection{Model Results}
\label{subsec:model_result}

We take the best fit of each model parameter directly from our Bayesian analysis, as the parameter value with the smallest $\chi^{2}$ in our multi-dimensional parameter space. For example, the SMBH mass is determined to be $5.75\times10^{8}$~$M_{\odot}$, consistent with the predicted value from the known stellar velocity dispersion ($216$~km~s$^{-1}$; \citealt{2013MNRAS.432.1709C}) and the $M_{\rm BH}$--$\sigma$ relation of \citet{2013ApJ...764..184M}.

The error bars of each model parameter are computed as percentiles of the posterior. Specifically, the $1\sigma$ confidence limits are the 15.9$^{\rm th}$ and  84.1$^{\rm th}$ percentiles of the posterior, while the 3sigma limits are the 0.14$^{\rm th}$ and 99.87$^{\rm th}$ percentiles of the posterior (see the grey-shaded regions in the likelihood contours of Figure~\ref{fig:histocontours}).

Using this procedure, the SMBH mass and stellar $M/L$ are measured to be $5.75^{+1.49}_{-1.18}\times10^{8}$~$M_{\odot}$ and $1.45\pm0.04$~$(M/L)_{\odot, H}$, respectively, at the $3\sigma$ confidence level. The reduced chi-square ($\chi^{2}_{\rm red}$) for the best fit is $0.75$, indicating a good fit. The inclination angle is measured to be $69\degr\!\!.90\pm0.61$ under the particular morphology of the molecular gas disc assumed. See Table~\ref{table:params} for the other best-fit parameters describing the molecular gas disc.

\begin{figure}
 \includegraphics[width=\columnwidth]{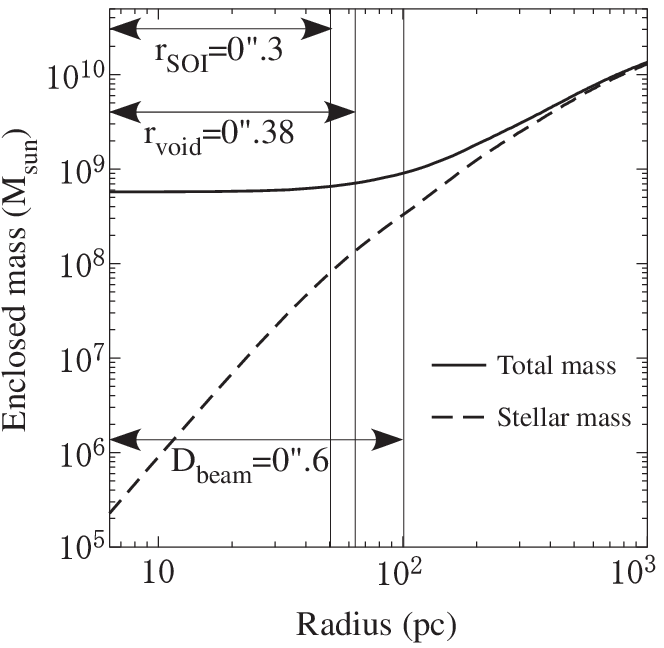}
  \caption[Enclosed Mass Along Radius]{An enclosed mass profile (solid curve) and an enclosed stellar mass profile (dashed curve) as considered in the MGE models.  Vertical lines show the SMBH sphere of influence (SOI) radius of $0\farcs3$, the radius of a central void ($0\farcs38$) derived from our fit, and the synthesized beam diameter ($0\farcs6$). The best-fit SMBH mass ($5.75\times10^{8}M_{\odot}$) is $6.98$, $3.73$ and $1.71$ times the enclosed stellar mass at each radius.}
  \label{fig:radstellarmass}
\end{figure}

This SMBH mass gives an intrinsic SOI of $0\farcs3$, half of the synthesized beam. This SOI radius is slightly smaller than the radius of the cavity in the best fit, $0\farcs38$. 
The SMBH mass is thus constrained to $23\%$ at the $3\sigma$ level ($7\%$ at $1\sigma$), although we do not detect Keplerian motion in the very centre.
Radial plot of the enclosed mass (Figure~\ref{fig:radstellarmass}) shows that the best-fit SMBH mass is $6.98$ times the enclosed stellar mass within $0\farcs3$ (SMBH SOI), $3.73$ times the stellar mass within $0\farcs38$ (molecular void radius) and $1.71$ times the stellar mass within $0\farcs6$ (synthesized beam size). An uncertainty for stellar mass within the SOI is less than $2.0\times10^{7}M_{\odot}$. We consider this to be negligible for the SMBH mass error.
We further investigate our SMBH mass error budget in Section~\ref{subsec:err}, focusing on possible systematic effects.

% Table: Model parameters
\begin{table*}
  \begin{center}
    \caption{Model parameters.}
    \label{table:params}
    \begin{tabular*}{\textwidth}{@{\extracolsep{\fill} }lcrcc} \hline
      Parameter & Search Range & Best Fit & Error ($1\sigma$ conf.) & Error ($3\sigma$ conf.) \\ 
      \hline
      SMBH mass ($10^{8}M_{\odot}$) & $0.01$--$50.12$ & $5.75$ & $+0.42, -0.38$ & $+1.49, -1.18$ \\
      Stellar $M/L$ ($M/L_{\odot, H}$) & $0.10$--$4.00$ & $1.45$ & $\pm0.01$ & $\pm0.04$ \\
      \\
      Molecular gas disc: \\
      Centre X offset (arcsec) & $-3.50$--$3.50$ & $-0.25$ & $\pm0.01$ & $\pm0.02$ \\
      Centre Y offset (arcsec) & $-3.50$--$3.50$ & $0.02$ & $\pm0.01$ & $\pm0.03$ \\
      Inclination ($\degr$) & $67.00$--$89.00$ & $69.90$ & $\pm0.20$ & $\pm0.61$ \\
      Position angle ($\degr$) & $0$--$50$ & $26\phantom{.00}$ & $\pm0$ & $\pm1$ \\
      Centre velocity offset (km~s$^{-1}$) & $-50.00$--$50.00$ & $-8.12$ & $\pm0.75$ & $\pm2.50$ \\
      Velocity dispersion (km~s$^{-1}$) & $1.00$--$20.00$ & $12.53$ & $\pm0.74$ & $\pm2.09$ \\
      Luminosity scaling & $10.00$--$200.00$ & $86.16$ & $\pm2.91$ & $\pm8.63$ \\
      Scale length (arcsec) & $1.00$--$7.00$ & $2.11$ & $\pm0.01$ & $\pm0.03$ \\
      Central void radius (arcsec) & $0.01$--$0.90$ & $0.38$ & $\pm0.03$ & $\pm0.04$ \\
      \hline
    \end{tabular*}
  \end{center}
  Notes: The prior distribution of each parameter, shown in the second column, is assumed to be uniform in linear space (logarithmic for the SMBH mass only). The posterior distribution of each parameter is quantified in the third to fifth columns (but see also Figure~\ref{fig:histocontours}). Please see text for the error estimation of $1\sigma$ ($68.3\%$) and $3\sigma$ ($99.7\%$) confidence. The central offset (X, Y) is an offset between the model and VLBI observations \citep[][]{2009A&A...505..509L}.
\end{table*}

% Figure: Enclosed Mass Plot

% Section --- Discussion
\section{Discussion}
\label{sec:dis}

% Subsection --- CO Moprhology
\subsection{CO Morphology}
\label{subsec:morph}

As NGC~3665 harbours an AGN, the molecular gas could be affected by its presence and/or that of the SMBH itself. With the highest angular resolution achieved so far, our CO observations have revealed that the molecular gas disc in NGC~3665 has an unresolved central hole, the position of which coincides with that of the radio core (detected with VLBI observations; \citealt{2009A&A...505..509L}) and the peak of the stellar surface brightness (identified from \textit{HST} observations).

The absence of molecular gas within a radius of $\approx0\farcs4$ or $\approx65$~pc suggests that some mechanism may be dissociating the molecular gas or preventing it from forming or accumulating in the very centre of the galaxy. The dissociation of molecules generally has two main causes, AGN activity and UV radiation from young stars. While the former is clearly a possibility, the latter is unlikely to be significant in NGC~3665 as the star formation density is low \citep{2014MNRAS.444.3427D}. Dynamical effects can also affect the distribution and survival of molecular gas. Shocks and resonances are obvious possibilities, but it may also be that the strong shear expected near the SMBH (where the circular velocity curve varies with the radius $R$ as $R^{-1/2}$) can destroy molecular clouds, where most molecules are generally found. Such dynamical mechanisms will be analysed further in future works. Evidently, the lack of molecular gas in the central hole also suggests that there is no current cold gas accretion in that region. Having said that, observations targeting different molecules and/or transitions are necessary to prove that the central hole is truly devoid of gas, and to explore possible mechanisms for void creation.

% Subsection --- AGN properties
\subsection{AGN Properties}
\label{subsec:AGN}

As mentioned above, given NGC~3665's stellar velocity dispersion \citep{2013MNRAS.432.1709C}, the best-fit SMBH mass is in agreement with the latest $M_{\rm BH}$--$\sigma$ relations \citep[e.g.][]{2013ApJ...764..184M}. However, we can also investigate whether this SMBH mass is consistent with the known properties of the AGN. A radio jet was detected early on in NGC~3665 \citep{1986A&AS...64..135P, 2016MNRAS.458.2221N}, with the jet axis almost exactly
perpendicular to the major axis of the central molecular gas disc. While AGN jets are generally thought to emerge perpendicularly 
to their accretion discs, it is also commonly accepted that there need not be a connection with the orientation of the large-scale disc of the galaxy. 
While the latter is the case for low-luminosity AGN such as Seyferts in late-type galaxies, the jet and large-scale molecular gas disc in NGC~3665 clearly have a connection. The kinematic position angle of the molecular gas disc is determined to be $26 \degr$, roughly perpendicular to the position angle of the jet, $137 \degr$ (determined by drawing a line from the northeast to the southwest blob seen at 5~GHz by \citealt{2016MNRAS.458.2221N}).

The X-ray luminosity of NGC~3665 ($L_{\rm X}$) was estimated to be $10^{40.1}$~erg~s$^{-1}$ in the $2$--$10$~keV energy range, extrapolated from the total energy within the \textit{Chandra} energy range of $0.3$--$8$~keV by assuming a power-law spectrum $N(E)=\alpha\,E^{1.7}$ and a value of $\alpha$ derived from the same observations \citep{2011ApJS..192...10L}. Comparing this X-ray luminosity to the Eddington luminosity calculated from our best-fit SMBH mass ($L_{\rm Edd}=1.27\times10^{38}\,M_{\rm BH}$~erg~s$^{-1}$~$M_{\odot}^{-1}=10^{46.8}$~erg~s$^{-1}$), we obtain an Eddington ratio of $\log(L_{\rm X}/L_{\rm Edd})=-6.73$. This relatively low Eddington ratio suggests radiatively inefficient flows, including powerful outflows such as the radio jet observed in NGC~3665 (\citealt{2003MNRAS.345.1057M} and references therein). We therefore conclude that the AGN properties of NGC~3665 do not conflict with the derived SMBH mass nor with the observed molecular gas kinematics.

% Subsection --- SMBH mass error
\subsection{Other Error Sources on the SMBH Mass}
\label{subsec:err}

Several high angular resolution observations of kinematics, aiming to measure SMBH masses, show Keplerian upturns in galaxy centres \citep[see, e.g., ][]{1995Natur.373..127M, 2016ApJ...822L..28B}. 
Keplerian motion is expected in a potential dominated by a SMBH, where observations resolve the SMBH SOI and some emission arises from within the SOI.
Spatially resolving the Keplerian region naturally allows to constrain the SMBH mass to high accuracy, as only a point mass yields such a behaviour, but our data do not show clear Keplerian motion.
A possible reason for this is that CO emission is not present in the SMBH's vicinity, as our model reveals a central cavity of radius $0\farcs38$ (see Section~\ref{subsec:model_result}). Even if that were not the case, the synthesized beam ($0\farcs60\times0\farcs56$; see Section~\ref{subsec:obsdr}) could have smeared out the information from within the SOI ($0\farcs3$; see Section~\ref{subsec:model_result}). 

Nevertheless, we measure a SMBH mass with an uncertainty of only $23\%$ (at $3\sigma$ confidence). The stellar $M/L$ and inclination, that have a direct influence on the SMBH mass, also have small uncertainties. Here we therefore investigate possible systematic errors on these parameters, and then discuss other possible effects that could increase the error budget. 

A potential reason for the small error on the stellar $M/L$ is the rather large fitting area, set to $13\farcs0 \times 13\farcs0$ so as to include all CO emission. A large fraction of this area is dominated by the stars rather than the SMBH. 
The uncertainty on the stellar $M/L$ (and in turn the SMBH mass and inclination) may thus decrease as the number of constraints (i.e. the area) increases, irrespective of  whether the fit is good or not, simply because the model looses its freedom to vary ($\chi^{2}$ would otherwise increase unacceptably).
We therefore narrow down the fitting area to $4\farcs0 \times 4\farcs0$, and repeat the fit described in Section~\ref{sec:model}. The SMBH mass is then measured to be $M_{\rm BH}=(5.37^{+1.24}_{-1.10})\times10^{8}$~$M_{\odot}$ and the stellar $M/L=1.49\pm0.05$~$(M/L)_{\odot, H}$, with an inclination angle $i=69\degr\!\!.64\pm0.77$, all at $3\sigma$ confidence level. This best-fit parameter set yields $\chi^{2}_{\rm red}=1.20$. Surprisingly, with fewer data, the constraints on the SMBH mass ($19\%$ error), $M/L$ ($3\%$ error) and inclination ($1\%$ error, all at $3\sigma$ confidence) are just as tight as the original result. Comparing these results with the ones from the original fit ($M_{\rm BH}=5.75^{+1.49}_{-1.18} \times 10^{8}$~$M_{\odot}$,  $(M/L)_{H}=1.45\pm0.04$~$(M/L)_{\odot, H}$ and $i=69\degr\!\!.90\pm0.61$, all at $3\sigma$; see Section~\ref{subsec:model_result}), we nevertheless notice a possible systematic error on the stellar $M/L$, but not on the SMBH mass or inclination. 

\begin{figure}
 \includegraphics[width=\columnwidth]{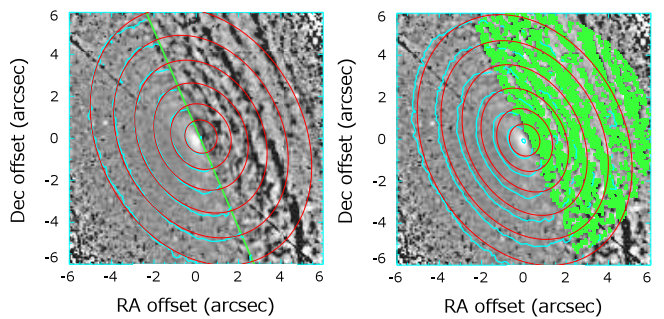}
  \caption[\textit{HST} $H$-band image and alternative MGE models.]{Left: Alternative MGE model (red contours) created by using data only on the left side of the green line. The contours are overlaid on an unsharp-masked \textit{HST} $H$-band (NICMOS F160W) image (blue contours and grey scale). Right: Another MGE model (red contours) created by masking all the pixels indicated in green, overlaid to the unsharp-masked \textit{HST} $H$-band (NICMOS F160W) image (blue contours and grey scale). }
  \label{fig:MGEalt}
\end{figure}

The systematic error on the stellar $M/L$ could come from the MGE model (stellar luminosity profile), that has no associated errors. The stellar luminosity profile becomes degenerate with the stellar $M/L$ when calculating the circular velocity, and thus clearly affects the uncertainty on the stellar $M/L$. Our MGE model also fluctuates depending on, for example, how we define the region of dust attenuation (see Section~\ref{subsec:model_lumdis} and Figure~\ref{fig:MGE} for the MGE model fitting). 
We thus use two more MGE models with different definitions of the region affected by dust, and fit to an area of $13\farcs0 \times 13\farcs0$. We first create a mask to cover the right half of the \textit{HST} image, divided with a line as shown in the left-hand panel of Figure~\ref{fig:MGEalt}. The unmasked region within a $6\farcs0$ radius is used to create the new MGE. The resulting SMBH mass is $M_{\rm BH}=5.4\times10^{8}$~$M_{\odot}$ with $(M/L)_{H}=1.41$~$(M/L)_{\odot, H}$, yielding $\chi^{2}_{\rm red}=0.73$. 
Second, we mask all the pixels with negative values in an unsharp-masked version of the {\textit HST} image, created using a Gaussian of FWHM $0\farcs2$. The mask and the unsharp-masked image are shown in the right-hand panel of Figure~\ref{fig:MGEalt}. By using the MGE fit created from the unmasked pixels, the SMBH mass and stellar $M/L$ are measured to be $M_{\rm BH}=5.50\times10^{8}$~$M_{\odot}$ and $(M/L)_{H}=1.24$~$(M/L)_{\odot, H}$, yielding again $\chi^{2}_{\rm red}=0.73$. The best-fit values of the stellar $M/L$ are however beyond the statistical $3\sigma$ error in both cases. 

% Figure: Alternative MGE model

Our tests thus indicate that systematic uncertainties on the SMBH mass are $\approx0.4\times10^{8}$~$M_{\odot}$, likely of the order of the quoted statistical $3\sigma$ errors.
The systematic uncertainties on the stellar $M/L$ are estimated to be $\approx0.2$~$(M/L)_{\odot, H}$, larger than the quoted statistical $3\sigma$ errors.

Here we do not explicitly consider the effects of potential spatial variations of the stellar $M/L$ onto the SMBH mass error budget. This could however lead to non-negligible uncertainties, and we plan to test for this in a future work. The uncertainties on the stellar $M/L$ and SMBH mass would also increase by considering e.g., galaxy distance and \textit{HST} photometric zero point uncertainties.

We further comment that the $\chi^{2}$ in our fit is calculated by considering only the rms noise level in the cube as the observational error. Including other sources of error would possibly increase the error on the fitting parameters.

% Subsection --- PVD fit
\subsection{Rotation Curve Fit}
\label{subsec:PVDfitting}

CARMA and ALMA observations just recently revealed that molecular gas kinematics enables SMBH mass measurements in nearby galaxies (e.g.\ NGC~4526, \citealt{2013Natur.494..328D}; NGC~1097, \citealt{2015ApJ...806...39O}; NGC~1332, \citealt{2016ApJ...823...51B}). However, when comparing to models, the first two studies only used a rotation curve extracted from a position-velocity diagram (PVD) taken along the galaxy major axis. The procedure can be briefly summarised: (1) determine the kinematic major axis and draw a PVD along it; (2) estimate the mean velocity at each position along the PVD to determine the rotation curve; (3) compare these velocity measurements with analogous ones made from a model rotation curve extracted from a model data cube (assuming a set of model parameters) in an identical manner; (4) identify the best-fit model parameters using a $\chi^{2}$ analysis throughout parameter space. 

In this sub-section, we thus re-derive the SMBH mass of NGC~3665 by fitting only the rotation curve extracted from our data, to verify that the value derived is in agreement with the full data cube fit of the previous sections. First, we fixed all molecular gas disc parameters to those obtained from the full data cube fit (see Table~\ref{table:params}), leaving only the SMBH mass and stellar $M/L$ as free model parameters. We then extracted the kinematic major-axis PVD from our data cube, and fit a Gaussian to the line-of-sight velocity distribution at each position to determine the mean velocity at that position, only keeping for the model fit measurements with a signal-to-noise ratio $S/N>3$. The uncertainty at each position was set to the root mean square of the channel width of the observation and the FWHM of the fitted Gaussian. If the Gaussian FWHM was too large, generally indicating a bad fit, we also excluded that position from the model fit. A total of $55$ mean velocity measurements were thus made along the extracted PVD, creating an observed rotation curve along the kinematic major-axis of NGC~3665.

\begin{figure}
  \includegraphics[width=\columnwidth]{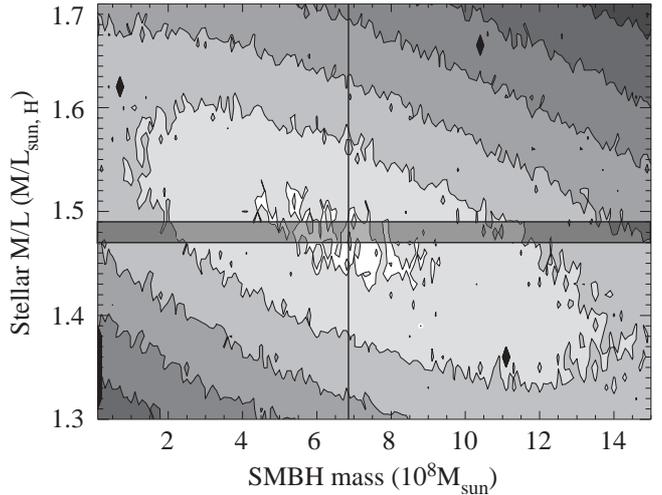}
  \caption[PVD fit uncertainties]{ $\chi^{2}$ contours in the parameter space of SMBH mass and stellar $M/L$. 
The contour levels are set to be $\chi^{2}_{\rm min}+53$, $212$, $477$, $848$ and $1325$. A grey shaded box and a black line indicate positions of cuts along each axes for the polynomial fits in Figure~\ref{fig:pvderr}.}
  \label{fig:pvdchiscont}
\end{figure}

\begin{figure}
  \includegraphics[width=\columnwidth]{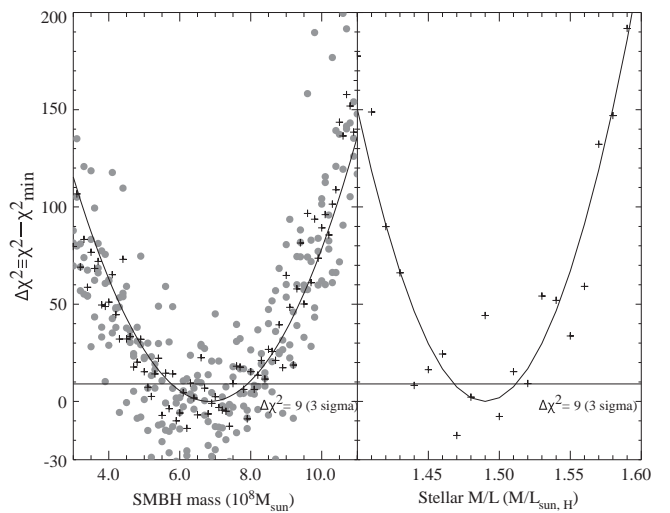}
  \caption[PVD fit uncertainties]{Distributions of the $\Delta\chi^{2}\equiv\chi^{2}-\chi^{2}_{\rm min}$ for the free parameters, SMBH mass (left) and stellar $M/L$ (right). 
Grey circles in the left panel are the actual $\Delta\chi^{2}$ achieved with the stellar $M/L=1.47$ to $1.49$ (from within the grey shaded area in Figure~\ref{fig:pvdchiscont}), and black crosses are values marginalized over the $M/L$. 
Right panel shows a polynomial fit to the distribution of $\Delta\chi^{2}$ at $M_{\rm BH}=6.8\times10^{8}M_{\odot}$ (cut along the vertical black line in Figure~\ref{fig:pvdchiscont}). 
The uncertainty of each parameter ($3\sigma$ confidence level) is determined by the intersection of the polynomial fits with the straight horizontal line ($\Delta\chi^{2}=9$). 
The best-fit model parameters are then $M_{\rm BH}=6.8^{+1.1}_{-1.0}\times10^{8}$~$M_{\odot}$ and $M/L=1.47\pm0.02$~$(M/L)_{\odot, H}$. }
  \label{fig:pvderr}
\end{figure}

Model data cubes are generated for a range of SMBH mass and $M/L$ values, and model rotation curves extracted in the same manner as for the data. These are then compared to the data in a $\chi^{2}$ manner.
Figure~\ref{fig:pvdchiscont} shows $\chi^{2}$ contours in the parameter space. By taking the average of upper and lower edge of the lowest contour level ($\chi^{2}_{\rm min}+53$), we first determine the stellar $M/L=1.48\pm0.01$~$(M/L)_{\odot, H}$ (grey shaded area in Figure~\ref{fig:pvdchiscont}) as a range to extract the $\chi^{2}$ distribution of the SMBH mass.
The $\chi^{2}$ values are then marginalized over the $M/L$ range, and a polynomial fit to the $\chi^{2}$ is determined.
$\Delta\chi^{2}\equiv\chi^{2}-\chi^{2}_{\rm min, fit}$ distribution of $M/L=1.48\pm0.01$~$(M/L)_{\odot, H}$ (grey circles), marginalized values (black crosses)
and the polynomial fit (black line) are shown in the left panel of Figure~\ref{fig:pvderr}. 
The minimum $\chi^{2}$ for the polynomial fit along the SMBH mass axis, thus $\chi{2}_{\rm min, fit}$, is realized with the SMBH mass of $6.8\times10^{8}M_{\odot}$. 
A polynomial fit to the $\chi^{2}$ distribution along the $M/L$ is then determined with the SMBH mass fixed to $6.8\times10^{8}M_{\odot}$ (along the vertical black line in Figure~\ref{fig:pvdchiscont}), as shown in the right panel of Figure~\ref{fig:pvderr}. 
The uncertainties on each parameter were estimated from the $99.7\%$ confidence interval (i.e.\ the parameter values with $\chi^{2}\leq \chi^{2}_{\rm min}+9$). The best-fit SMBH mass and stellar $M/L$ are then $6.8^{+1.1}_{-1.0}\times10^{8}$~$M_{\odot}$ and $1.47 \pm 0.02$~$(M/L)_{\odot, H}$, respectively, with a $\chi^{2}_{\rm min}$ of $129.8$ and $\chi^{2}_{\rm red}=2.45$ (the number of degrees of freedom here is $53$, allowing for the two free parameters).  
We note that the minimum $\chi^{2}_{\rm red}$ is above unity, suggesting the existence of a better fit with other molecular gas disc parameters.
A scatter seen in the $\chi^{2}$ distributions (see Figure ~\ref{fig:pvderr}) can produce a systematic error for each parameters, but we do not further discuss about this here. We do not expect the systematic error of SMBH mass to be larger than $0.4\times10^{8}M_{\odot}$ and $M/L$ to be larger than $0.2$~$(M/L)_{\odot, H}$, both of which is derived in Section~\ref{subsec:err}. We stress that the SMBH SOI is not resolved in our observations, and the Keplarian motion is not seen in our PVD. Resolving the SMBH SOI can narrow down the error budget of the SMBH mass in this rotation curve analysis.

\begin{figure*}
   \includegraphics[width=17.0cm]{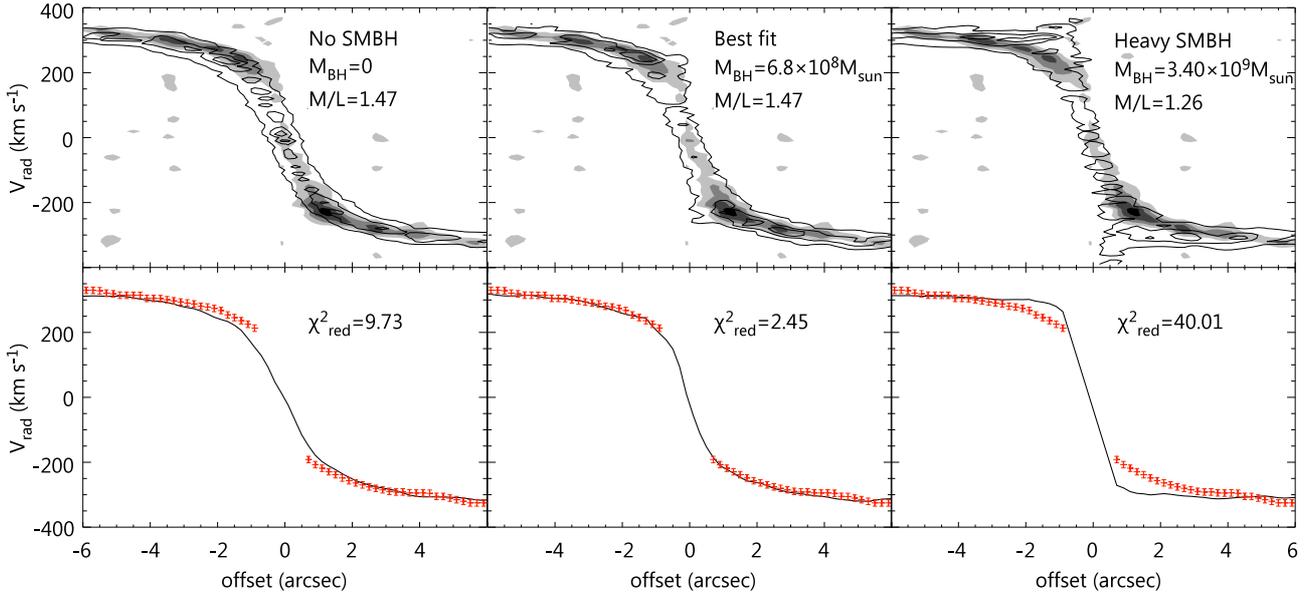}
  \caption[Comparison of position-velocity diagrams from model and
  observation]{Upper panels: Observed position-velocity diagram (PVD) along the kinematic major axis (greyscale), overlaid with the model PVDs (contours). The left panel shows a comparison with a model without a black hole ($M_{\rm BH}=0$), the middle panel with the best fit, and the right panel with an overweight black hole.
    Lower panels: Extracted mean velocities along the kinematic major axis (red points with error bars), overlaid with the best-fit model velocities (black lines). The reduced $\chi^{2}$ of each model is $9.73$ (left, $M_{\rm BH}=0$), $2.45$ (middle, best fit) and $40.01$ (right, overweight SMBH). }
  \label{fig:pvdfitting}
\end{figure*}

A comparison of the observed and best-fit model PVD and their associated rotation curves is also shown in Figure~\ref{fig:pvdfitting}. The two parameters for each panel are  $M_{\rm BH}=0$ with a best-fit $M/L=1.48$~$(M/L)_{\odot, H}$ (left panel; $\chi^{2}_{\rm red}=9.73$), the best-fit $M_{\rm BH}=6.8\times10^{8}$~$M_{\odot}$ and $M/L=1.47$~$(M/L)_{\odot, H}$ (middle panel; $\chi^{2}_{\rm red}=2.44$) and an overweight SMBH with $M_{\rm BH}=3.4\times10^{9}$~$M_{\odot}$ and $M/L=1.26$~$(M/L)_{\odot, H}$ (right panel; $\chi^{2}_{\rm red}=40.01$).

% Figure: PVD fit uncertainties -- contours

% Figure: PVD fit uncertainties --polynomial fits

% Figure: PVD fit

The SMBH mass and stellar $M/L$ derived from the rotation curve fit are both consistent with those obtained from the full data cube fit, despite the very small uncertainties on the $M/L$ ratios. In fact, the uncertainties derived are also very similar. The two fitting methods therefore appear equally useful a priori. 
This is probably because the rotation curve fit extracts the crucial part of the data cube, most closely related to the mass of the SMBH and stellar body. However, we recall that the rotation curve fit has only $2$ free parameters, compared to $11$ for the data cube fit. 
The other parameters, taken from the data cube fit, were fixed during the rotation curve fit. The data cube thus contains a lot of information useful to constrain many other model parameters associated with the structure of the molecular gas disc itself (see Section~\ref{sec:model}).
In addition, a major disadvantage of the rotation curve fit is that some parameters can be degenerate with the SMBH mass or stellar $M/L$, and therefore must be independently constrained (e.g.\ the disc inclination, that directly affect the $M/L$ and to a lesser extent the SMBH mass). Extracting the appropriate PVD also requires a well-defined kinematic major axis, that is difficult to specify when the velocity field shows warps and/or kinematic twists. A middle ground is to fit an extracted velocity field. For example, \citet{2007ApJ...671.1329N} considered warped and misaligned ellipses to model the ionised gas velocity field of NGC~5128 (Centaurus~A). In NGC~3665, however, there is no significant evidence for non-circular motions. Models with warped structures may nevertheless be required in the future, as highly detailed gas distributions gradually become available through higher angular resolution observations.

%
% Section --- Conclusions
%
\section{Conclusions}
\label{sec:conc}

We present CARMA $^{12}$CO$(J=2-1)$ observations of the early-type galaxy NGC~3665 with $0\farcs59$ resolution. These reveal a regularly rotating molecular gas disc in the equatorial plane of the galaxy, with an apparent void within a radius of $\approx0\farcs4$ or $\approx65$~pc, potentially created by the known AGN.

Fitting the entire observed data cube of NGC~3665 with a model with free SMBH mass, stellar $M/L$, and numerous parameters describing the structure of the molecular gas disc, we derive a SMBH mass of $M_{\rm BH}=5.75^{+1.49}_{-1.18}\times10^{8}$~$M_{\odot}$ and a stellar $M/L$ of $(M/L)_{H}=1.45\pm0.04$~$(M/L)_{\odot, H}$ at $3\sigma$ confidence levels (statistical error). 
The SOI of the SMBH is thus estimated to be $0\farcs3$, which is half of the synthesized beam. The central hole in the gas disc ($0\farcs38$) limits our gas tracer to radii outside the SOI, and all of the dynamical constraints on the SMBH mass come from model fitting to data outside of the SOI, where stellar mass is dominant.
This SMBH mass is in agreement with that estimated from the latest $M_{\rm BH}-\sigma$ correlations, and appears consistent with the known AGN properties of NGC~3665, such as its radio jet and X-ray luminosity. Systematic uncertainties on the stellar $M/L$ are estimated to be $\approx0.2$~$(M/L)_{\odot, H}$, by considering smaller fitting regions and several different MGE models. We estimate the systematic uncertainties on the SMBH mass to be $\approx0.4\times10^{8}$~$M_{\odot}$, which is within the statistical $3\sigma$ error.
The full data cube fit also yields a SMBH mass consistent with that derived from a fit to the rotation curve only, but it opens the door to SMBH mass measurements in sources with significantly more complex molecular gas discs. 

This work is only the fourth SMBH mass measurement using molecular gas kinematics, following measurements in two other lenticular galaxies and one barred spiral \citep{2013Natur.494..328D, 2015ApJ...806...39O, 2016ApJ...823...51B}. This method has thus now proven its usefulness to derive SMBH masses in various types of galaxies. It offers exciting prospects to both calibrate SMBH masses measured with other methods, and to simply increase the number of galaxies with reliable SMBH masses. Further investigations comparing SMBH masses measured using other methods (stellar kinematics, ionised-gas kinematics and/or megamasers) will be required before a proper discussion of potential systematic differences between the different methods is possible.

%
% Section --- Acknowledgements
%
\section*{Acknowledgements}
% Referee
The authors thank the referee, Aaron Barth, for constructive comments and useful suggestions.
%My fund
Part of this study was financially supported by JSPS KAKENHI grant number 26*368.
KO acknowledges support from a Royal Astronomical Society.
MB was supported by STFC consolidated grant `Astrophysics at Oxford' ST/H002456/1 and ST/K00106X/1. 
MC acknowledges support from a Royal Society University Research Fellowship.
TAD acknowledges support from a Science and Technology Facilities Council Ernest Rutherford Fellowship.
%CARMA obs. acknowledgement
Support for CARMA construction was derived from the states of California, Illinois and Maryland, the James S.\ McDonnell Foundation, the Gordon and Betty Moore Foundation, the Kenneth T.\ and Eileen L.\ Norris Foundation, the University of Chicago, the Associates of the California Institute of Technology and the National Science Foundation. CARMA science was supported by the National Science Foundation under a cooperative agreement, and by the CARMA partner universities.
%IRSA obs.
This research made use of the NASA/IPAC Infrared Science Archive, which is operated by the Jet Propulsion Laboratory, California Institute of Technology, under contract with the National Aeronautics and Space Administration.
%HST obs.
This work is also based on observations made with the NASA/ESA \textit{HST}, obtained from the Data Archive at the Space Telescope Science Institute, which is operated by the Association of Universities for Research in Astronomy, Inc., under NASA contract NAS 5-26555.
%Data Analysis
Data analysis was partially carried out on the common use data analysis computer system at the Astronomy Data Center, ADC, of the National Astronomical Observatory of Japan.

%%%%%%%%%%%%%%%%%%%%%%%%%%%%%%%%%%%%%%%%%%%%%%%%%%


%%%%%%%%%%%%%%%%%%%%%%%%%%%%%%%%%%%%%%%%%%%%%%%%%%

%%%%%%%%%%%%%%%%% APPENDICES %%%%%%%%%%%%%%%%%%%%%

%\appendix


%%%%%%%%%%%%%%%%%%%%%%%%%%%%%%%%%%%%%%%%%%%%%%%%%%

% Don't change these lines
\bsp	% typesetting comment
\label{lastpage}
\end{document}